\documentclass{ws-ijmpa}
\usepackage{amsmath}
\usepackage{graphicx}
\def\ast{\alpha^*}

\def\pdellx'{\frac{\partial}{\partial x'}}
\def\pdellw'{\frac{\partial}{\partial w'}}

\newcommand{\be}{\begin{equation}}
\newcommand{\ee}{\end{equation}}
\def\bed{\begin{displaymath}}
\def\eed{\end{displaymath}}
\def\bea{\begin{eqnarray}}
\def\eea{\end{eqncrray}}
\def\[{$$}
\def\]{$$}
\begin{document}

\title{Description and prediction of even-A nuclear masses based on residual proton-neutron interactions}

\author{Bao-Bao JIAO	\footnote{E-mail:baobaojiao91@126.com}\\
	Department of Physics, University of Shanghai for Science and Technology\\Shanghai 200093, People's Republic of China\\}

\date{Jul 16, 2018}

\maketitle
{\small The odd-even staggering of neighboring nuclear masses is very useful in
calculating local mass relations and nucleon-pair correlations.
During the past decades, there has been an increasing interest in the odd-even features of the mass relations and related quantities exhibited in masses of neighboring nuclei.
In this work, after choosing a nucleus, we made an analysis of its neighboring nuclei on the upper left corner and the lower right corner respectively.
We empirically obtained a new residual interaction formula of even-$A$  ($A$ is the mass number) nuclei,
and it is an revision based on the existing empirical local formula of the proton-neutron interactions between
the last proton and the last neutron ($\delta V_{1p-1n}$).
We then calculated the even-$A$ nuclear masses. The differences between our calculated values
and the AME2012 database show that the root-mean-squared deviations (RMSD) are small (for even-$A$ nuclei: $A$ $\geq$ 42, RMSD $\approx$ 161 keV; $A$ $\geq$ 100, RMSD $\approx$ 125 keV), while for heavy nuclei,
some of our calculated values can reach an accuracy of a few tens of keV.
With our residual interaction formula including one parameter, we have successfully predicted some unknown masses.
Some of our predicted values compared well with the experimental values (AME2016).
In addition, the accuracy and simplicity of our predicted masses
for medium and heavy nuclei are comparable to those of the AME2012 (AME2016) extrapolations.}

keywords: Residual proton-neutron interactions; nuclear masses; binding energies.


\section{Introduction}

Nuclear masses$^{1-16}$
and energy levels are important issues in the field of nuclear physics.
For a given nucleus with $Z$ protons and $N$ neutrons, the relationship between binding energy $B(Z,N)$
\cite{lab17,lab18,lab19,lab20}
and nuclear mass $M(Z,N)$ ($B(Z,N) = ZM_p + NM_n - M(Z,N)$,
where the $M_p$ and the $M_n$ are the mass of a free proton and a free neutron)
is of great importance in areas of physics,
such as nuclear structure and fundamental interactions.
The binding energy and the nuclear mass are also useful for nuclear astrophysics.
The atomic mass evaluation (AME) was published in 2012 and 2017 (AME2012 \cite{lab21} and AME2016 \cite{lab22}),
in which approximately two hundred additional nuclei were listed than in AME2003 \cite{lab23}.

The description and evaluation of the nuclear masses are one of the focuses in nuclear structure physics.
There is a significant quantity of research in this direction.
In nuclear physics, there are many mass models and mass formulas.
Generally, mass formulas are divided into two major categories: global mass relations and local mass relations.
The first one is global mass relations. Earlier studies are as follows: the famous Weizs$\ddot{a}$cker
formula \cite{lab1} and the finite range droplet model \cite{lab2}.
The BCS theory based on the relativistic mean field model \cite{lab3,lab4} has reached
an accuracy of root-mean-squared deviation RMSD $\approx$ 500 keV, the Skyrme-Hartree-Fock-Bogoliubov
theory \cite{lab5,lab6} RMSD $\approx$ 561 keV, the finite range droplet model \cite{lab2,lab7} RMSD $\approx$ 570 keV, the Duflo-Zuker model \cite{lab10}.
The second is local mass relations.
Local mass relations have also proved to be useful for the application of Coulomb displacement energies of mirror nuclei in mass predictions.
Such as Audi-Wapstra systematics, the Garvey-Kelson (G-K) mass relations \cite{lab11}
(for even-$A$ nuclei with $A$ $\geq$ 100, RMSD $\approx$ 170 keV), the nucleon-pair correlations mass
relations $^{24-30}$.
There are two empirical formulas (after choosing a nucleus, they made an analysis of its neighboring nuclei on the lower left corner), even-A nuclei formula and odd-A nuclei formula respectively in Ref. 24.

Our purpose in this paper is to describe a new $\delta V_{1p-1n}$ local formula that can be useful in describing and predicting even-$A$ nuclear masses.
After choosing a nucleus ($M(Z,N)$), we made an analysis of its neighboring nuclei on the upper left corner $(M(Z+1,N)$, $M(Z,N-1)$, $M(Z+1,N-1))$ and the lower right corner ($M(Z-1,N+1)$, $M(Z,N+1)$, $M(Z-1,N)$) respectively.
We obtained a new local formula of even-A nuclei based on the empirical formula in Ref. 24.
There are comparatively good agreement between the calculated and experimental values
(the RMSD between our calculated values and the AME2012 database: for even-$A$ nuclei with $A$ $\geq$ 42, RMSD $\approx$ 161 keV, $A$ $\geq$ 100, RMSD $\approx$ 125 keV;
the RMSD between our calculated values and the AME2016 database: for even-$A$ nuclei with $A$ $\geq$ 42, RMSD $\approx$ 164 keV, $A$ $\geq$ 100, RMSD $\approx$ 126 keV;
the RMSD between our calculated values and the AME2003 database: for even-$A$ nuclei with $A$ $\geq$ 42, RMSD $\approx$ 181 keV, $A$ $\geq$ 100, RMSD $\approx$ 134 keV),
while for medium-mass and heavy nuclei, the calculated values are in good agreement with the AME databases.
The study of residual interaction is helpful in describing experimental values,
which demonstrates that our local formula can be used to predict unknown masses.
The focus is that we can use one parameter of the proton-neutron interactions
formula to describe and predict the even-$A$ nuclear masses.

The structure of the paper is as follows.
Section 2 briefly reviews the new residual interaction formula of even-A nuclei.
We then discuss the RMSDs of known even-$A$ nuclear masses.
In addition, our manuscript gives an estimate of extrapolation uncertainties.
In section 3, by applying experimental values (AME2012) and our new local formula,
we successfully predict some unknown masses and discuss their deviations.
We note that our predicted masses are very close to those predicted in the AME2012 database,
in particular, those with A $\geq$ 42.
The result demonstrates that some of our predicted values and the
experimental values in AME2016 \cite{lab22} have good accuracy and agree well.
In section 4, we discuss and summarize the results of our work.

\section{Residual proton-neutron interactions}

Residual proton-neutron interactions play an important role in nuclear physics.
For the past few years, they have attracted more and more attention $^{31-44}$.
The study of proton-neutron interactions is very helpful in studying shell model theory and phase transitions
$^{45-47}$.
We obtained a new $\delta V_{1p-1n}$ local formula of even-A nuclei based on the
empirical formula (in Ref. 24) and the neighboring nuclei (neighboring nuclei on the upper
left corner and the lower right corner).
Then, we use our empirical local formula to describe and predict the even-$A$ nuclear masses.
After choosing a nucleus, we made an analysis of its neighboring
nuclei on the lower right corner. The residual interaction between the last proton and the last neutron is
\begin{eqnarray}
\label{eq1}
\delta V_{1p-1n}(Z,N+1)=B(Z,N+1)+B(Z-1,N)-\nonumber\\[1mm]
\ B(Z,N)-B(Z-1,N+1)\nonumber\\[1mm]
\ =M(Z,N)+M(Z-1,N+1)-\nonumber\\[1mm]
\ M(Z,N+1)-M(Z-1,N).
\end{eqnarray}

\subsection{Residual proton-neutron interactions}
We empirically obtained the $\delta V_{1p-1n}(Z,N+1)$ formula of even-$A$ (where $A=Z+N$) nuclei based on the above study.
We successfully describe and predict some even-$A$ nuclear masses from
experimentally known nuclear masses and the $\delta V_{1p-1n}$ formula.

\begin{figure}[!htb]
	\label{fig1}
	\centering
	\includegraphics[scale=0.3]{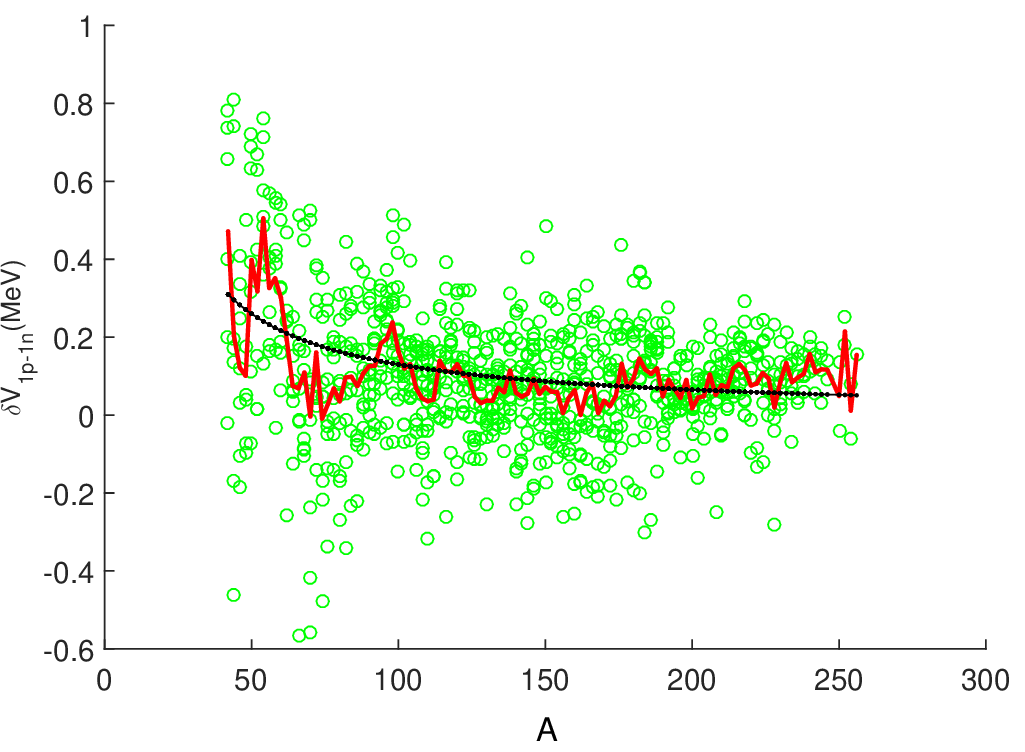}
	\caption{(color online) Circles show that the $\delta V_{1p-1n}$ for $A \geq$ 42 ($A$ is even). The red zig zag line is plotted by using the average values of $\delta V_{ip-jn}$ for nuclei with the same mass number, expressed as $\overline{\delta V_{1p-1n}}$.
		The black curve is plotted in terms of equation (6): $\overline{\delta V_{1p-1n}}(A) = \frac{13000}{A}$ keV .}
	\label{placas_paralelas}
\end{figure}

We obtained about 920 groups of datasets(for A $\geq$ 42) based on Eq. (1).
Then we use a definition of the binding energy B(Z,N) which yields a positive quantity and obtains a positive $\delta V_{1p-1n}$.
However, after choosing a nucleus, in Ref. 24 the authors made an analysis of its
neighboring nuclei on the lower left corner, they obtained the $\delta V_{1p-1n}$  formulas of odd-A and even-A ($A\geq100$) nuclei:
\begin{eqnarray}
\label{eq2}
B(Z,N)+B(Z-1,N-1)-B(Z,N-1)-B(Z-1,N) \nonumber\\[1mm]
\ \approx\overline{\delta V_{1p-1n}}(A) \approx -74 \mathrm{keV}, \nonumber\\[1mm]
B(Z,N)+B(Z-1,N-1)-B(Z,N-1)-B(Z-1,N) \nonumber\\[1mm]
\ \approx\overline{\delta V_{1p-1n}}(A) \approx -74-\frac{69861}{A} \mathrm{keV}.
\end{eqnarray}

As shown by using the $\overline{\delta V_{1p-1n}}(A+1)$, we empirically have

\begin{eqnarray}
\label{eq3}
\overline{\delta V_{1p-1n}}(A+1) \simeq B(Z,N+1)+B(Z-1,N)-B(Z,N)\nonumber\\[1mm]
\ -B(Z-1,N+1) \simeq \frac{13000}{A}
\ \mathrm{keV}.
\end{eqnarray}
where $A$ is even and $A \geq$ 42.
This formula is the focus of discussion in our paper.
Our empirical local formula approximately reflected the features of proton-neutron interactions.
The main advantage of our local formula given in Eq. (3) is that it involves masses of only four neighboring nuclei and one parameter.
They are important for reliable predictions in the process of iterative extrapolations.
The smaller the number of nuclei involved in local mass relations,
the more reliable the predictions in iterative extrapolations,
and the smaller the deviations are in the extrapolation process, see Ref. 48.

\subsection{RMSDs and uncertainties}
Based on the experimental masses and $\delta V_{1p-1n}$, we get the binding-energy formula and the mass equation:
\begin{eqnarray}
\label{eq4}
B(Z,N)=B(Z,N+1)+B(Z-1,N)-\nonumber\\[1mm]
\ B(Z-1,N+1)-\overline{\delta V_{1p-1n}}(A+1).
\end{eqnarray}
\begin{eqnarray}
\label{eq5}
M(Z,N)=M(Z,N+1)+M(Z-1,N)-\nonumber\\[1mm]
\ M(Z-1,N+1)+\overline{\delta V_{1p-1n}}(A+1).
\end{eqnarray}

The root-mean-square deviation (RMSD) of the masses is defined as usual:
\begin{eqnarray}
\label{eq11}
\sigma = \sqrt{\frac{1}{n}\sum^{n}_{i=1}(M_i^{exp}-M_i^{cal})^2}.
\end{eqnarray}
Using the local mass relation (Eq. (5)), we obtained even-A nuclear masses($M^{cal}$) by
combining our new empirical local formula with three experimental values (AME2012).
We use our one-parameter empirical local formula to describe and predict the nuclear residual interaction, then to describe and predict the even-A nuclear masses.
Equation (6) reflects the error of our empirical local formula (3) for the residual interaction.
We obtained the RMSDs based on the method in Refs. [24,49],
and then compared our precision to other local mass relations.
The result demonstrates that the RMSD of our local mass relation is a little
smaller than other local mass relations.
The RMSD of even-A nuclei is presented in Fig. 2.
We obtained even-$A$ nuclear masses by experimental values and our local formula: $\overline{\delta V_{1p-1n}}(A+1) \simeq \frac{13000}{A}$ $\mathrm{keV}$.
Solid triangles in black are plotted by using the RMSDs ($\sigma_{2012}$) between our calculated values and the AME2012 databases. Five-pointed stars in red correspond to the RMSDs ($\sigma_{2016}$) between our calculated values and the AME2016 database.

\begin{figure}[!htb]
	\label{fig2}
	\centering
	\includegraphics[scale=0.5]{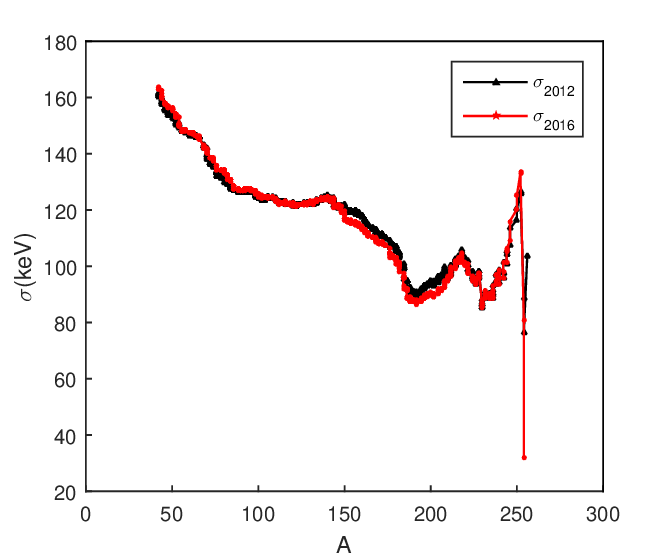}
	\caption{(color online) Root-mean-squared deviations of nuclear masses for $A\geq42$.}
	\label{placas_paralelas}
\end{figure}

As is depicted in Fig. 2, the differences show that RMSDs are small,
while for heavy nuclei,
our calculated values can reach an accuracy of below 100 keV (for even-$A$ nuclei with $A$ $\geq$ 192, RMSD $\approx$ 90 keV).
In fact, this is already shown in Fig. 1, where result for heavy nuclei are more precise than light nuclei.
The result proves that our calculated values have good accuracy and compared well with AME databases (AME2012 and AME2016).

The RMSD between our calculated values and the AME2003: for even-$A$ nuclei with $A$ $\geq$ 42, RMSD $\approx$ 181 keV; $A$ $\geq$ 100, RMSD $\approx$ 134 keV.
The RMSD in Ref. 24: compared with AME2003, for even-A nuclei with $A$ $\geq$ 100, RMSD $\approx$ 168 keV; for odd-A nuclei with $A$ $\geq$ 100,
RMSD $\approx$ 132 keV.
In Refs. [24,50], the difference of $\delta V_{1p-1n}$ between odd-A nuclei and their neighboring even-A nuclei (A$\geq$100) is $\frac{69900}{A}$ keV, we empirically have:
\begin{eqnarray}
\label{eq12}
\overline{\delta V_{1p-1n}}(A+1) \simeq B(Z,N+1)+B(Z-1,N)-B(Z,N)\nonumber\\[1mm]
\ -B(Z-1,N+1) \simeq \frac{82900}{A}
\ \mathrm{keV}.
\end{eqnarray}
We obtained the odd-A nuclear masses through Eq. (7).
For odd-$A$ nuclei ($A$ $>$ 100):
compared with AME2012, RMSD $\approx$ 151 keV; compared with AME2016, RMSD $\approx$ 153 keV; compared with AME2003, RMSD $\approx$ 155 keV.
For even-A nuclei, our new empirical local formula is more precise than the formula in Ref. 24.
But our paper lacks precise empirical local formula for odd-A nuclei.
The focus of Eq. (3) is that it involves parameter of only one,
while the number of parameters involved in Ref. 24 is two.
The main advantage of our empirical local formula has good accuracy and compared well with AME databases (AME2003, AME2012 and AME2016).
Our calculated values confirm that our new formula can be used to predict unknown even-A nuclei masses.

Additionally, we use the procedure described by Refs. [2,24] to "decouple" the experimental errors. The procedure is based on maximum-likelihood method, we using MATLAB program to estimate the extrapolation uncertainties.
The iteration algorithm can be written as in Algorithm 1 (in the Appendix).
Numerical experiments show that the $\sigma_{th}$ is 153 keV for $A\geq42$ (For $A$ is even and $A\geq100$: $\sigma_{th} \approx$ 113 keV ).
We take the same $\sigma_{th}$ in predicting in the same mass region based on Fig. 3.
Now we evaluate uncertainties for unknown masses in our predictions by Eq. (3),
so the uncertainty ($\sigma_{pred}$) of our predicted value is given by:
\begin{eqnarray}
\label{eq13}
[\sigma_{pred}(Z,N)]^2=[\sigma_{th}(A)]^2+[\sigma_{exp}(Z,N+1)]^2+\nonumber\\[1mm]
[\sigma_{exp}(Z-1,N)]^2+[\sigma_{exp}B(Z-1,N+1)]^2.
\end{eqnarray}

\begin{figure}[!htb]
	\label{fig3}
	\centering
	\includegraphics[scale=0.3]{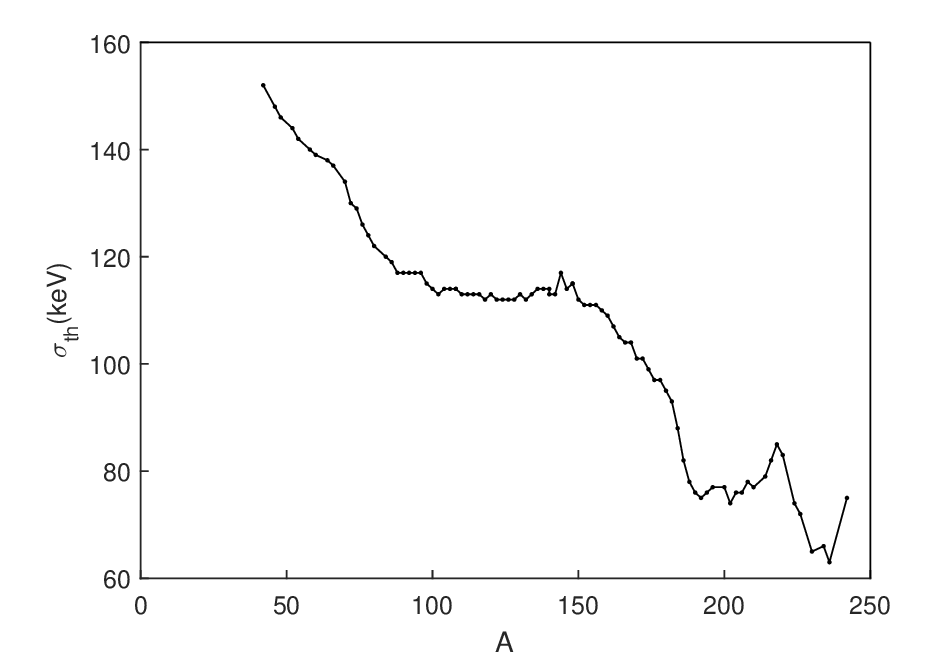}
	\caption{(color online) The curve is plotted by fitting these calculated theoretical errors ($\sigma_{th}$).}
	\label{placas_paralelas}
\end{figure}

\section{Prediction of nuclear masses}

In this section, by combining our new empirical local formula with AME2012, we successfully predict some nuclear masses which are not experimentally accessible.
We predict unknown masses with Eq. (5).
At the same time, the average binding energies of our predicted masses are in good agreement with the
curve of specific binding energy \cite{lab51}.
We predicted the residual proton-neutron interactions and binding energies of unknown masses based on Eqs. (3) and (4);
then we can get the mass excess (ME$^{pred}$).
In table 1 we present a set of selected data of our predicted values and predicted values in AME database (AME2003 (ME$^{2003}$) and AME2012 (ME$^{2012}$)).
Obviously, our predicted values have good accuracy and compared well with AME2012 database.
Some nuclei that are important either in astrophysics or in nuclear structure.

\begin{table}[h]
\tbl{Mass excess of predicted. (keV)}
{\begin{tabular}{@{}cccccccc@{}} \toprule
Nucleus  & ME$^{2003}$ & ME$^{2012}$ & ME$^{pred}$ & Nucleus  & ME$^{2003}$ & ME$^{2012}$ & ME$^{pred}$\\
\colrule
$^{48}$Fe	&	-18160	&	-18416	&	-17722	&	$^{178}$Tl	&	-4750	&	-4794	&	-4674	 \\
$^{50}$Co	&	-17200	&	-17782	&	-17344	&	$^{190}$Tl	&	-24330	&	-24379	&	-24470	 \\
$^{52}$Co	&	-33920	&	-33990	&	-34279	&	$^{190}$At	&	Null	&	Null	&	7074	 \\
$^{54}$Cu	&	-21690	&	-21741	&	-21803	&	$^{192}$Rn	&	Null	&	Null	&	10039	 \\
$^{60}$Ga	&	-40000	&	-39784	&	-39959	&	$^{198}$At	&	-6670	&	-6721	&	-6643	 \\
$^{62}$Mn	&	-48040	&	-48481	&	-48238	&	$^{198}$Fr	&	Null	&	Null	&	9533	 \\
$^{62}$Ge	&	-42240	&	-41899	&	-41860	&	$^{202}$Fr	&	3140	&	3092	&	3142	 \\
$^{64}$As	&	-39520	&	-39652	&	-39340	&	$^{206}$Ac	&	13510	&	13462	&	13482	 \\
$^{66}$Se	&	-41720	&	-41368	&	-41295	&	$^{220}$Pa	&	20380	&	20218	&	20239	 \\
$^{68}$Br	&	-38640	&	-38441	&	-38311	&	$^{222}$Pa	&	22120	&	22155	&	22115	 \\
$^{70}$Kr	&	-41680	&	-40948	&	-40827	&	$^{224}$Np	&	Null	&	31876	&	31771	 \\
$^{82}$Zr	&	-64190	&	-63943	&	-63401	&	$^{232}$Np	&	37360	&	37361	&	37203	 \\
$^{84}$Nb	&	-61880	&	-61021	&	-60615	&	$^{234}$Am	&	44530	&	44461	&	44384	 \\
$^{114}$I	&	-72800	&	-72796	&	-72682	&	$^{246}$Es	&	67900	&	67902	&	67926	 \\
$^{150}$Tm	&	-46610	&	-46491	&	-46601	&	$^{256}$Md	&	87620	&	87457	&	87362	 \\
$^{152}$Lu	&	-33420	&	-33422	&	-33559	&	$^{258}$Db	&	101750	&	101799	&	101567	 \\
\botrule
\end{tabular}}
\end{table}

Now let us focus on a few examples of our predicted values.
Table 1 shows that $^{190}$At, $^{192}$Rn and $^{198}$Fr
are not predicted in the AME2003 and AME2012 databases.
Very interestingly, for $^{198}$Fr ($^{206}$Ac) the deviation of our predicted masses from the experimental results \cite{lab22} is only $\sim$ 37 keV (2 keV).
The additional nuclei is $^{224}$Np,
its value cannot be predicted in the AME2003 database,
and the difference between our predicted value and predicted value (Ref. 21) is approximately 100 keV.

\begin{table}[pt]
\tbl{The difference between the predicted and experimental values(AME2016). (keV)}
{\begin{tabular}{@{}cccccccccccc@{}} \toprule
Nucleus     & Exp. & $\delta$0  & ME$^{pred}$ & $\delta$1 & $\delta$2 &Nucleus      & Exp. & $\delta$0  & ME$^{pred}$ & $\delta$1 & $\delta$2\\
\colrule
$^{52}$Co 	&	-34361	&	8	&	-34279	&	144	&	-82  	&	$^{190}$Tl  	&	-24372	&	 8	&	 -24470	&	84	&	98	\\
$^{56}$Cu 	&	-38643	&	15	&	-38504	&	141	&	-139	&	$^{194}$Bi  	&	-16029	&	 6	&	 -15946	&	93	&	-83	\\
$^{62}$Mn 	&	-48524	&	7	&	-48238	&	241	&	-286	&	$^{198}$Fr  	&	9570	&	 30	&	 9533	&	95	&	37	\\
$^{82}$Zr  	&	-63631	&	11	&	-63401	&	121	&	-230	&	$^{202}$Fr  	&	3096	&	 7	&	 3142	&	91	&	-46	\\
$^{84}$Nb 	&	-61219	&	13	&	-60615	&	120	&	-604	&	$^{206}$Ac  	&	13480	&	 50	&	 13482	&	117	&	-2	\\
\botrule
\end{tabular}}
\end{table}

Table 2 shows that our predicted values (ME$^{pred}$) and the experimental values (Exp.) in AME2016 coincide well. We use $\delta$0 to represent the experimental deviations.
The uncertainty ($\sigma^{pred}$) of our predicted values is given by $\delta$1.
The $\delta$2 corresponds to deviations between our calculated values and the experimental values.
It is easy to observe that a comparison of our predicted values and experimental values shows that the deviation is a few hundreds of keV.
Some of our predicted values can reach an accuracy of a few tens of keV or several keV.
The systematicness of the residual interaction is better in the heavy nuclei than in the light nuclei region, which leads to the large deviation in the light nucleus region.
Moreover, we empirically obtained the residual proton-neutron interactions formula by using $\overline{\delta V_{1p-1n}}$, this is also a cause of the deviation ($^{84}$Nb).
The experimental values given in AME2016 are rounded.

Table 1 and Table 2 are studied based on AME2012 database, Table 3 is predicted by using AME2016 database. In Table 3, we present a set of selected data among our predicted values, which have a large difference between $\delta$3 and $\delta$4.
The $\delta$3 corresponds to deviations between ME$^{2016}$(predicted values in AME2016) and ME$^{pred1}$(our predicted values based on AME2012), and $\delta$4 represents the differences between ME$^{2016}$ and ME$^{pred2}$(our predicted values based on AME2016).
The $\delta$3 and $\delta$4 demonstrate that the predicted values obtained using accurate experimental values (AME2016) are closer to the new predicted values in AME2016.
In addition, we obtained some new predicted values (e.g, $^{196}$Fr, $^{200}$Ra, $^{204}$Ac, $^{214}$U).
More accurate predictions could be readily made if the odd-even features were more accurate \cite{lab52}.
Our new empirically formula has good accuracy and compared well with
proton-neutron interactions, therefore the local mass relation can be used to describe and predict the even-A nuclei masses (AME2003, AME2012 and AME2016).
 This work may be helpful for experimenters.

\begin{table}[h]
\tbl{The predicted values  based on AME2016. (keV)}
{\begin{tabular}{@{}ccccccc@{}} \toprule
Nucleus     & ME$^{2016}$   & ME$^{pred1}$ & ME$^{pred2}$ & $\delta$3 &$\delta$4 \\
\colrule
$^{52}$Ni	&	-22330	&	Null	&	-22362	&	Null	&	32		    \\
$^{54}$Cu	&	-21410	&	-21803	&	-21747	&	393	    &	337      	\\
$^{68}$Br	&	-38790	&	-38311	&	-38459	&	-479	&	-331		\\
$^{70}$Kr	&	-41100	&	-40827	&	-40975	&	-273	&	-125		\\
$^{78}$Y	&	-52170	&	-52820	&	-52281	&	650	    &	111		    \\
$^{114}$I	&	-72800	&	-72682	&	-72629	&	-118	&	-171		\\
$^{136}$Eu	&	-56240	&	Null	&	-56097	&	Null	&	-143		\\
$^{174}$Au	&	-14240	&	-13965	&	-14022	&	-275	&	-218		\\
$^{178}$Ta	&	-50600	&	-50732	&	-50630	&	132	    &	30		    \\
$^{178}$Tl	&	-4790	&	-4674	&	-4726.5	&	-116	&	-63.5		\\
$^{196}$Fr	&	Null	&	Null	&	13400	&	Null	&	Null		\\
$^{200}$Ra	&	Null	&	Null	&	12641	&	Null	&	Null		\\
$^{204}$Ac	&	Null	&	Null	&	16775	&	Null	&	Null		\\
$^{214}$U	&	Null	&	Null	&	25161	&	Null	&	Null		\\
$^{222}$Pa	&	22160	&	22115	&	22058	&	45	    &	102		   \\
$^{232}$Np	&	37360	&	37203	&	37285	&	157  	&	75		   \\
$^{258}$No	&	91480	&	Null	&	91435	&	Null	&	45		   \\
\botrule
\end{tabular}}
\end{table}

\section{Discussion and Conclusions}
In this work,
after choosing a nucleus, we made an analysis of its neighboring nuclei (neighboring nuclei on the lower right corner),
we obtained a new empirical local formula based on the empirical formulas in Ref. 24, then to describe and predict the residual proton-neutron interactions.
The result shows that our new local formula is useful to single mass extrapolation.

We study nuclear masses origin of the odd-even difference in terms of residual proton-neutron interactions $\delta V_{1p-1n}$.
We find a useful local formula based on $\delta V_{1p-1n}$ for even-$A$ nuclei with $A$ $\geq$42:
$B(Z,N+1)+B(Z-1,N)-B(Z,N)-B(Z-1,N+1) \simeq \frac{13000}{A}$ $\mathrm{keV}$.
We then obtained the RMSDs by comparing the calculative values with the experimental values
(compared with AME2012, for even-$A$ nuclei: $A$ $\geq$ 42, RMSD $\approx$ 161 keV; $A$ $\geq$ 100, RMSD $\approx$ 125 keV;
compared with AME2016, for even-$A$ nuclei: $A$ $\geq$ 42, RMSD $\approx$ 164 keV; $A$ $\geq$ 100, RMSD $\approx$ 126 keV;
compared with AME2003, for even-$A$ nuclei: $A$ $\geq$ 42, RMSD $\approx$ 181 keV; $A$ $\geq$ 100, RMSD $\approx$ 134 keV),
for the medium-mass and heavy nuclei, calculated values can reach an accuracy of a few tens of keV.
Comparing our predicted values with the AME2012 database shows that the deviations are small.
Our accurate and simple predictions of masses for medium and heavy
nuclei are comparable with those of the AME2012 extrapolations.
Additionally, some of our predicted values and experimental values (AME2016) are in good agreement.
These results show that our new residual interaction formula has good accuracy in describing and predicting the proton-neutron interactions, and our calculated values compared well with AME databases (AME2003, AME2012 and AME2016).

Our purpose here is to describe a new empirical residual proton-neutron interactions
formula that can be useful in describing and predicting masses of even-$A$ nuclei.
In predicting the unknown masses, our mass relation requires three nuclei.
The smaller the number of nuclei involved in local mass relations,
the more reliable the predictions in iterative extrapolations,
and the smaller the deviations are in the extrapolation \cite{lab48} process.
In addition, our residual proton-neutron interactions formula included one parameter.
The simple formula can help us understand the residual interaction.
This is another advantage of our mass relation.
We study the residual proton-neutron interactions and make use of these results in
evaluating and predicting the masses (AME2003, AME2012 and AME2016).
Further, more accurate predictions could be readily made if the predicted proton-neutron interactions were more accurate.

\appendix

\section{Appendices}
Assumed the theoretical errors are Gaussian-type distributed with centered at a mean value($\mu_{th}$) and
a standard deviation($\sigma_{th}$).
The algorithm has a high convergence speed therefore the maximum number of iterations is less than 20($j$) times.
Where $i$ is an abbreviation of $(Z,N)$, $w_i$ shows that the weight factor of the $ith$ nucleus.
Predicted and experimental values of binding energy of the $i$th nucleus defined as $B_{th}^i$ and $B_{exp}^i$.
The $\sigma_{exp}$ is denoted that the deviation of the binding energy involved in the prediction.
Here $(\alpha_i)^2$ denotes as: $(\alpha_i)^2(Z,N)=(\sigma_{exp})^2(Z,N+1)+(\sigma_{exp})^2(Z-1,N)+(\sigma_{exp})^2B(Z-1,N+1)$.\\

\noindent Algorithm 1 (maximum-likelihood method)

\noindent Input: $B_{th}^i$, $B_{exp}^i$, $\alpha_i$ and $\sigma_{exp}^i$.

\noindent Output: $\sigma_{th}$(convergence value)

set $\sigma_{th}$=100 (assume an initial value of $\sigma_{th}$);

for $j$=1:20

$w=\frac{1}{(\alpha_i)^2+(\sigma_{th})^2}$ (a series of weighting factors: $w=[w_1;w_2;...;w_k]$); \\

$\mu_{th}=\frac{\sum_{i=1}^kw_i(B_{exp}^i-B_{th}^i)}{\sum_{i=1}^kw_i}$ (where $i=1,2,...,k$; $k=920$);\\

$(\sigma_{th})^2=\frac{\sum_{i=1}^kw_i^2[(B_{exp}^i-B_{th}^i-\mu_{th})^2-(\sigma_{exp}^i)^2]}{\sum_{i=1}^kw_i^2}$ (with iteration); \\

if $\mid\sigma_{th}(j)-\sigma_{th}(j-1)\mid\leq10^{-4}$;

    break

end

end\\
The iteration of $\sigma_{th}$ was continued until the value converged. However, the convergence is found to be extremely rapid ($\sim$ 5 times).

\section*{Acknowledgements}
The author would like to thank L. Y. Jia for reading and commenting of this paper.
Support is acknowledged from the National Natural Science Foundation of China, Grant No. 11405109.

\end{document}